\documentclass[psfig]{aastex}

\usepackage{emulateapj5}

\def\ltsima{$\; \buildrel < \over \sim \;$}

\def\lsim{\lower.5ex\hbox{\ltsima}}
\def\gtsima{$\; \buildrel > \over \sim \;$}
\def\gsim{\lower.5ex\hbox{\gtsima}}

\begin{document}
\title{Detection of Gravitational Waves from the
Coalescence of Population-III Remnants 
with Advanced LIGO}

\author{J. Stuart B. Wyithe\altaffilmark{1} and Abraham
Loeb\altaffilmark{2}}

\email{swyithe@isis.ph.unimelb.edu.au; aloeb@cfa.harvard.edu}

\altaffiltext{1}{University of Melbourne, Parkville, Victoria, Australia}

\altaffiltext{2}{Harvard-Smithsonian Center for Astrophysics, 60 Garden
St., Cambridge, MA 02138}

\begin{abstract}
\noindent 
The comoving mass density of massive black hole (MBH) remnants from
pre-galactic star formation could have been similar in magnitude to the
mass-density of supermassive black holes (SMBHs) in the present-day
universe. We show that the fraction of MBHs that coalesce during the
assembly of SMBHs can be extracted from the rate of ring-down gravitational
waves that are detectable by Advanced LIGO. Based on the SMBH formation
history inferred from the evolution of the quasar luminosity function, we
show that an observed event rate of 1 yr$^{-1}$ will constrain the SMBH
mass fraction that was contributed by MBHs coalescence down to a level of
$\sim 10^{-6}$ for $20M_\odot$ MBH remnants (or $\sim 10^{-4}$ for
$260M_\odot$ remnants).
\end{abstract}


\section{Introduction}

The first episode of star formation (Pop-III) occurred in metal-free gas
that cooled to low temperatures through collisional excitation of molecular
hydrogen, H$_2$ (Bromm \& Larson 2003).  The first stars could have formed
inside mini-halos of masses $M\ga 10^5M_\odot$
at redshifts $z\ga 20$ (Barkana \& Loeb 2001). Numerical simulations
indicate that the first stars
were probably more massive than the sun by two orders
of magnitude (Bromm, Coppi \& Larson~1999; Abel, Bryan \&
Norman~2000). Pop-III stars with a mass $M_\star\ga 260M_\odot$ or $25\la
(M_\star/M_\odot) \la 140$ end their life by forming a MBH remnant (Heger
\& Woosley~2002; Heger et al. 2003).

The formation of Pop-III stars must have been self-regulating, as
the UV radiation emitted by these stars dissociated H$_2$ throughout the
universe (Haiman, Rees \& Loeb 1997) and boiled (via photo-ionization
heating) gas out of shallow potential wells with a virial temperature $\la
10^4$ K (Barkana \& Loeb 1999), whereas supernovae explosions expelled gas
from even deeper potential wells (Yoshida et al. 2003). Moreover, Pop-III
supernovae enriched star-forming regions with metals above the critical
value of $\sim 0.1\%$ of the solar metallicity, allowing fragmentation of
the gas into lower-mass stars (Bromm \& Loeb 2003).
Even with this self-regulation included, Madau \& Rees~(2001) estimate
that the co-moving mass density in MBHs at $z\sim 20$ may have been
comparable to the observed mass density of supermassive black holes (SMBHs)
in the present-day universe. Thus, in principle, the coalescence of MBH
remnants could have made a significant contribution to the SMBH mass
budget.

More recent semi-analytic calculations of the merger history of SMBHs
(Islam, Taylor \& Silk~2002) find that
although the mass density in Pop-III MBH remnants may be
sufficient to provide the present day density of SMBHs, 
most of the MBHs end up in satellites and populate galaxy halos.
Nevertheless, hierarchical merging inevitably leads to the formation of
early dwarf galaxies, which could sink
via dynamical friction to make dense clusters of MBHs at the centers of
massive galaxies that are assembled later (Madau \& Rees 2001). These
central MBH clusters provide sites for MBH coalescence and could produce
seeds for SMBH growth (analogous processes take place within the segregated
cores of globular star clusters; see Portegies-Zwart \& McMillan 2000).

The rise of the SMBH population is traced by the evolution of the quasar
luminosity function (e.g. Yu \& Tremaine~2002).  While the gas accretion
history of the quasar population accounts for most of the locally observed
SMBH mass, coalesced MBHs may also contribute some small fraction, $f_{\rm
MBH}$, of that mass.  For a given value of $f_{\rm MBH}$, the coalescence
rate of MBHs corresponding to the buildup of mass in SMBHs may then be
calculated based on the observed quasar history.  In this {\it Letter} we
show that detection of gravitational waves from MBH coalescences by {\it
Advanced LIGO}\footnote{http://www.ligo.caltech.edu/advLIGO/} (LIGO-II)
will constrain the fraction of the SMBH mass that was assembled through
coalescence of MBHs and provide a direct probe of the early population of
very massive stars.

\section{Ring-Down Gravitational Waves from MBHs and Detection by LIGO-II}
\label{ringdown}

First, we consider the expected gravitational radiation signal from merging
MBHs at high redshifts. The coalescence process of two black-holes can be
separated into three phases: an inspiral phase, a coalescence phase, and
finally a ringing phase of the merged black-hole product (Flanagan \&
Hughes~1998) as it settles to a Kerr metric. Of these, the inspiral and
ringing phases generate well understood wave-forms. Inspiraling MBHs at
high redshift will be most luminous at frequencies near 1 Hz, just below
the sensitivity band for LIGO-II. However, following a merger the
MBHs would ring at higher frequencies that are within the LIGO-II
sensitivity band.

\begin{figure*}[htbp]
\epsscale{1.4}
\plotone{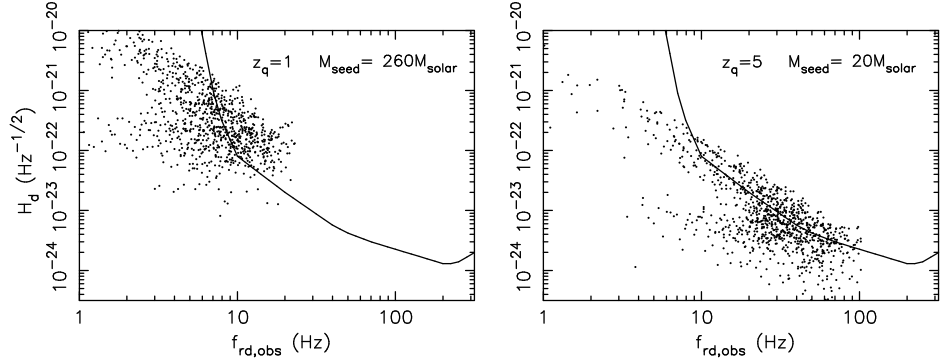}
\caption{\label{fig1} Comparison between the ring-down strain signals
$H_{\rm d}$ and the LIGO-II sensitivity curve as a function of observed
frequency, $f_{\rm rd,obs}$.  
The points correspond to hierarchical coalescence of $10^3$ MBH seeds to a
single MBH at a redshift $z_{\rm q}$.  Two cases are shown, having MBH
seeds of $20M_\odot$ and $260M_\odot$, coalescing by $z_{\rm q}=5$ and
$z_{\rm q}=1$, respectively.  Each coalescence event involves a different
randomly-chosen combination of the parameters $\theta$, $\phi$, $\psi$,
$\lambda$ and $\hat{L}\cdot\hat{n}$. The calculations assume $\epsilon_{\rm
rd}=0.03$ and $a=1.0$.}
\end{figure*}
To compute the gravitational wave strain for black-hole ringing we follow
the formalism outlined by Hughes~(2002).  The dimensionless strains for the
two polarizations of the gravitational waveform during the ring down may be
written as
\begin{equation}
\nonumber h_{+,-}(t) = \mathcal{A}_{+,-}\exp{(-\pi f_{\rm rd,obs}t/Q)}
\cos{(2\pi f_{\rm rd,obs}t)},
\end{equation}
where $Q\equiv\pi f_{\rm rd,obs}\tau_{\rm obs}$ is the quality
factor, $\tau_{\rm obs}$ is the observed decay time and $t$ is the observed
time following the merger.  The observed frequency, $f_{\rm rd,obs}$, 
equals the intrinsic frequency, $f_{\rm rd}$, divided by
$(1+z)$. Assuming that the ratio of the strain amplitudes for the two
polarizations follows that of the inspiral phase,
\begin{equation}
\mathcal{A}_+=\mathcal{A}_{\rm
ring}\left[1+(\hat{L}\cdot\hat{n})^2\right]\hspace{2mm}\mbox{;}\hspace{2mm}\mathcal{A}_\times=-2\mathcal{A}_{\rm
ring}(\hat{L}\cdot\hat{n}),
\end{equation}
where $\hat{L}\cdot\hat{n}$ is the dot product of the unit vectors pointing
along the binary orbital angular momentum and the line of sight to the
source. The overall strain amplitude $\mathcal{A}_{\rm rd}$ is found
by requiring that the binary radiates a fraction $\epsilon_{\rm rd}$ of its energy
(e.g. Fryer, Holz \& Hughes~2002). 
If a black hole of mass $\Delta M_{\rm bh}$ merges with a black hole of
mass $M_{\rm bh}$
\begin{equation}
\mathcal{A}_{\rm rd} =
10^{-23}\left(\frac{D}{\mbox{10~{\rm Gpc}}}\right)^{-1}\sqrt{\frac{20\epsilon_{\rm
rd}}{4\pi f_{\rm rd}Q}\left(\frac{M_{\rm bh}+\Delta M_{\rm
bh}}{M_\odot}\right)}
\end{equation}
Note that in this expression $D=D_{\rm L}/(1+z)$ is the co-moving distance,
where $D_{\rm L}$ is the luminosity distance (e.g. Hogg~1999). The energy
flux is proportional to $(f_{\rm rd,obs})^2 h_{\rm
ring}^2\propto \left[(f_{\rm rd})^2(1+z)^{-2}\right] D^{-2}\propto D_{\rm
L}^{-2}$.

Distortions of the Kerr metric during the ringing phase may be decomposed
into spheroidal modes with spherical harmonic like indicies $l$ and $m$
(Fryer, Holtz \& Hughes~2002). The quadrupole ($l=2$) moments are expected
to dominate ringdown spectrum. Binary coalescence excites a bar-like
deformation of the event horizon (Hughes~2002).  As a result, the $m=2$
mode dominates over the $m=0$ mode when the ringdown forms the final
stage of a binary merger.  Fits to numerical simulation show that for
$l,m=2$ (Leaver~1985; Echeverria~1989; Fryer, Holtz \& Hughes~2002)
\begin{equation}
\label{frequencies}
f_{\rm rd}
\sim \frac{10^{5.3}\mbox{Hz}}{2\pi}\left(\frac{M_{\rm bh}+\Delta M_{\rm
bh}}{M_\odot}\right)^{-1}\left[1-0.63(1-a)^{3/10}\right],
\end{equation}
and 
\begin{equation}
\nonumber
Q = 2(1-a)^{-9/20},
\end{equation}
where $a$ is the dimensionless black-hole spin, taken to have
representative values of $1$, $0.5$ or $0$ in the numerical results
presented later. Hughes \& Blandford~(2002) have found that equal mass
mergers may have rapidly rotating remnants ($a\sim1$), while mass ratios of
$0.5$ lead to typical spins of $a\sim0.5$. We note that in addition to
setting the oscillation frequency, the spin governs the value of $Q$ and
hence the number of oscillations. Ringdown waveforms with more oscillations
will be easier to detect.

The observed strain may be written as
\begin{equation}
H(t)=F_+(\theta,\phi,\psi)h_+(t) + F_\times(\theta,\phi,\psi)h_\times(t),
\end{equation}
where $F_+(\theta,\phi,\psi)$ and $F_+(\theta,\phi,\psi)$ are the detector
response functions
for a source with a sky position $(\theta,\phi)$ and polarization axes
rotated at an angle $\psi$ (e.g. Thorne~1987).  Since the ring-down waves
lie in a narrow frequency range, the signal-to-noise ratio $\rho$ for a
detection utilizing matched-filtering techniques may be evaluated as
(Hughes~2002)
\begin{equation}
\label{SNR}
\rho^2=\frac{2\int_{0}^\infty dt H(t)^2}{S_{\rm h}(f_{\rm rd,obs})},
\end{equation}
where $S_{\rm h}(f_{\rm rd,obs})$ is the spectral density of detector noise
(in Hz$^{-1}$).

\section{MBH mergers and the Evolution of the SMBH population}

In the redshift interval between $z_{\rm q}$ and $z_{\rm q}+dz_{\rm q}$, the accretion due to
quasars implies that the co-moving mass density in SMBHs is
\begin{equation}
\label{rho}
d\rho_{\rm bh}(z_{\rm q})=dz_{\rm q}\int_0^\infty dL_{\rm B} \frac{L_{\rm
bol}}{\epsilon c^2}\Psi(L_{\rm B},z_{\rm q})\frac{dt}{dz_{\rm q}},
\end{equation}
where $L_{\rm bol}$ is the bolometric luminosity of a quasar with a B-band
luminosity $L_{\rm B}$, $\epsilon$ is the conversion efficiency of accreted
mass to electromagnetic radiation (we adopt $\epsilon=0.1$; Yu \&
Tremaine~2001), and $\Psi(L_{\rm B},z_{\rm q})$ is the co-moving density of
quasars per unit B-band luminosity at $z_{\rm q}$, for which we adopt the
parametric form derived from the 2dF survey (Boyle et al.~2000).  By
integrating equation~(\ref{rho}) over redshift, one may obtain the SMBH
mass-density accreted by a redshift $z_{\rm q}$. The local ($z_{\rm q}=0$)
value for the co-moving mass density in SMBHs accreted during the quasar
activity is $\rho_{\rm bh}\approx 2.7\times10^5M_\odot\mbox{Mpc}^{-3}$.
This number is similar to estimates of the local SMBH mass density obtained
from the inventory of quiescent galactic SMBHs (e.g. Yu \& Tremaine~2002;
Aller \& Richstone~2002).

If some fraction of the SMBH mass is assembled through coalescence during
the hierarchical build-up of galaxies, then it is natural to expect MBH
remnants to contribute to the SMBH mass budget as they are more massive
than normal stellar remnants and should sink to the centers of galaxies by
dynamical friction (Madau \& Rees 2001).  The initial coalescence of the
MBHs would be detectable by Advanced LIGO, and the observed event rate (or
lack thereof) could shed light on the possibility that MBH coalescence
events produced the early seeds of SMBHs.  Suppose that a fraction $f_{\rm
MBH}$ of the mass in SMBHs at any redshift was added through coalescence of
MBHs, with the remainder added through gas accretion.  Since the comoving
mass density of accreted gas mass for quasars is comparable to the final
SMBH mass density in the local universe, we expect $f_{\rm MBH}\ll 1$
independent of other considerations.

While the accreted mass is being added at redshift
$z_{\rm q}$, the coalescence of the MBHs incorporated into the SMBH at
$z_{\rm q}$ would have stretched over most of the age of the universe
$t_{\rm age}(z_{\rm q})$, corresponding to the assembly time of the SMBH.
The MBHs are brought together through hierarchical merging of the
mini-halos that form the building blocks for the massive dark matter halo
hosting any particular SMBH. There are several issues to consider,
including the number and clustering of MBHs inside mini halos, the merger
rate of mini-halos into proto-galaxies and more massive hosts, and the
subsequent dynamical behavior of the MBHs within the resulting gaseous and
stellar environments. Rather than model the coalescence history using a
dark-matter halo merger tree and the poorly understood dynamics of the MBH
population, we instead assume that the MBHs undergo random hierarchical
coalescences with the MBH mergers distributed linearly in time. 
At any point in
the coalescence history, two MBHs that coalesce are chosen at random from
the MBH population.  The next merger is then drawn from the revised
population which contains one fewer MBH in total, and one additional MBH of
a larger mass, and so on. There are a total of $N-1$ coalescence events
leading from the $N$ initial seed MBHs to the single final MBH of mass
$NM_{\rm MBH}$. One third of these coalescence events are between MBHs of
the initial seed mass.

\begin{figure*}[htbp]
\epsscale{1.4}
\plotone{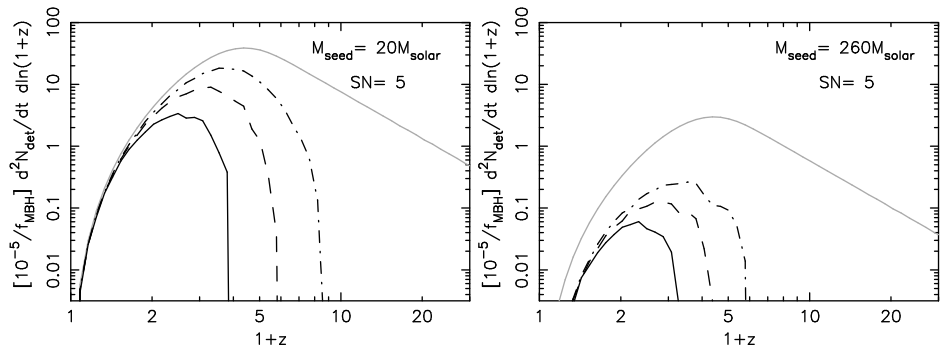}
\caption{\label{fig2}The rate of detectable mergers per  $\ln(1+z)$
in units of $10^5 f_{\rm MBH}~{\rm yr}^{-1}$.
Curves are shown for a signal-to-noise ratio of 5, and initial MBH mass of
$20M_\odot$ (left panels) and $260M_\odot$ (right panels). Each panel shows
three curves for $\epsilon_{\rm rd}=0.01$ (solid), 0.03 (dashed) and 0.10
(dot-dashed).
The calculations assumed $a=1.0$. 
For comparison, we also show the total number of mergers per year in the
observer frame (light lines).}
\end{figure*}

Figure~\ref{fig1} shows the signal amplitude $H_{\rm
d}\equiv\sqrt{2\int_{0}^\infty dt H(t)^2}$ for the hierarchical coalescence
of $10^3$ MBH seeds to a single MBH at a redshift $z_{\rm q}$. Two cases
are illustrated, having MBH seeds of $20M_\odot$ and $260M_\odot$
coalescing by $z_{\rm q}=5$ and $z_{\rm q}=1$, respectively.  Each
individual coalescence event was combined with a different randomly chosen
combination of the parameters $\theta$, $\phi$, $\psi$, $\lambda$ and
$\hat{L}\cdot\hat{n}$. The calculations assume $\epsilon_{\rm rd}\sim0.03$
(Hughes~2002), and $a=1.0$.  The values of $H_{\rm d}$ in Figure~\ref{fig1}
lie along two branches.  The lower branch originates from the early stage
of the merger tree, where the seed MBHs provide most of the coalescence
events. As the typical mass of the coalescing MBHs grows at lower
redshifts, the signal amplitude increases while its frequency
decreases. The figure also shows $\sqrt{S_{\rm h}(f_{\rm rd,obs})}$, 
the planned sensitivity curve for the two LIGO-II 4km
interferometers (Gustafson, Shoemaker, Strain \& Weiss~1999). The
signal-to-noise ratio $\rho$ is the ratio between the points and the
sensitivity curve [see equation~(\ref{SNR})]. From an ensemble of such
detection histories, we may therefore compute the fraction of mergers
$f_{\rm detect}(z_{\rm q},z_{\rm coal})$ that are detectable by LIGO-II as
a function of the quasar redshift, $z_{\rm q}$, and the coalescence
redshift, $z_{\rm coal}$.

\begin{table*}[tbp]
\begin{center}
\caption{\label{tab1} Detection rate of coalescence events in units of
$10^5f_{\rm MBH}$yr$^{-1}$. 
The third column shows the merger rate assuming that all events can be
detected.}
\begin{small}
\begin{tabular}{cccccc}
\hline 
                &     &                                                & \multicolumn{3}{c}{$[10^{-5}~{\rm yr}/f_{\rm MBH}]\times dN_{\rm det}/dt$} \\
 $M_{\rm seed}$ & $a$ & $[10^{-5}/f_{\rm MBH}]\times dN_{\rm mrg}/dt$  & $\epsilon_{\rm rd}=0.01$  & $\epsilon_{\rm rd}=0.03$  & $\epsilon_{\rm rd}=0.10$  \\\hline
$20M_\odot$ & 1.0 &  35.6& 1.9  & 5.7  &  13.3  \\   
$20M_\odot$ & 0.5 &  35.6& 1.0   & 3.4   & 9.8     \\   
$20M_\odot$ & 0.0 &  35.6& 0.67  & 2.3   & 7.5     \\   
$260M_\odot$ & 1.0 &  2.7  & 0.02 & 0.07  & 0.15 \\ 
$260M_\odot$ & 0.5 &  2.7  & 0.003 & 0.01   & 0.03   \\ 
$260M_\odot$ & 0.0 &  2.7  & 0.001 & 0.003  & 0.007  \\ \hline
\end{tabular}
\end{small}
\end{center}
\end{table*}

The rate at which coalesced MBH mass is added to the SMBH population is given by
\begin{equation}
\frac{d\rho_{\rm MBH}}{dz_{\rm q}} = \frac{f_{\rm MBH}}{1-f_{\rm
MBH}}\frac{d\rho_{\rm bh}}{dz_{\rm q}},
\end{equation}
with the remainder added via gas accretion during the active quasar phase.
The number density increment of consumed MBHs can be derived from the mass
density increment, $dn_{\rm MBH}(z_{\rm q})=d\rho_{\rm MBH}(z_{\rm
q})/M_{\rm MBH}$. 

The redshift distribution of the rate of detectable coalescences (in the
observer frame) is
\begin{eqnarray}
 \nonumber \frac{d^2N_{\rm det}}{dtdz_{\rm coal}} &=& \int_\infty^0 dz_{\rm
q} \left[dn_{\rm MBH} 4\pi\frac{dV}{d\Omega dz_{\rm q}} \frac{f_{\rm detect}(z_{\rm q},z_{\rm coal})}{[t_{\rm
age}(z_{\rm q})]^2} \right.\\ &&\hspace{5mm}\left. \frac{dt}{dz_{\rm
coal}} \left(\frac{1}{1+z_{\rm
coal}}\right)\Theta(z_{\rm coal}-z_{\rm q})\right],
\end{eqnarray}
where the factor ${1}/{(1+z_{\rm coal})}$ is due to cosmological time
dilation, the derivative ${dt}/{dz_{\rm coal}}$ enforces a linear
distribution of coalescence events in time, $({dV}/{d\Omega dz_{\rm q}})$
is the co-moving volume per unit solid angle per unit redshift, and $\Theta
(x)$ is the Heaviside step function.

Figure~\ref{fig2} shows ${d^2N_{\rm det}}/{dtd\ln(1+z_{\rm coal})}$ in
units of $10^5 f_{\rm MBH}~{\rm yr^{-1}}$.
The plotted curves correspond to a signal-to-noise ratio of 5, assuming a
MBH initial mass of $20M_\odot$ (left panels) and $260M_\odot$ (right
panels).  The calculations were performed for values of $a=1.0$ and
$\epsilon_{\rm rd}=0.01$, 0.03 and 0.10.  A fraction as small as $f_{\rm
MBH}\sim10^{-5}$ yields $\sim0.1$--$10$ detectable coalescence events at
redshifts $2\la z_{\rm q}\la 5$ (where the quasar population is substantial
and the observed frequency does not redshift out of the Advanced LIGO band
above $10$ Hz).  For comparison, the light lines show the redshift
distribution of the entire coalescence rate ($f_{\rm detect}=1$) in the
observer frame, ${d^2N_{\rm mrg}}/{dtd\ln(1+z_{\rm coal})}$.  We find that
Advanced LIGO will be able to detect most of the low redshift coalescences,
while the detectable fraction declines at early cosmic times due to the
redshift in frequency.

Values of the total event rate $({dN_{\rm det}}/{dt})$ are presented in
Table~\ref{tab1} in units of $10^5f_{\rm MBH}~{\rm yr^{-1}}$. In addition
to the examples plotted in Figure~\ref{fig2}, the table lists results for
BH spins of $a=0.0$ and $a=0.5$. The rates are lower for smaller MBH spins
and for larger MBH masses, due to the lower emission frequency and smaller
numbers of MBHs.

\section{Discussion}
\label{discussion}

We have estimated the number of seed MBHs that must have coalesced at high
redshifts in order to contribute a fraction $f_{\rm MBH}$ of the mass
density in quasar SMBHs.  We then found the corresponding detection rate of
ring-down gravitational waves for the Advanced LIGO interferometers. If the
initial mass of MBH remnants was $20M_\odot$, then 1 event per year would
imply a coalesced mass fraction of $f_{\rm MBH}\sim10^{-6}$. On the other
hand, if the MBH seed mass was $260M_\odot$, then 1 event per year would
imply a coalesced mass fraction of $f_{\rm MBH}\sim10^{-4}$.  Advanced LIGO
will be primarily sensitive to MBH coalescence at $z\sim2-3$, while our
calculation implies that most of the MBHs for which coalescence is possible
may have coalesced earlier. However, values of $f_{\rm
MBH}\sim10^{-4}-10^{-6}$ correspond to only a few events per typical
galactic nucleus per Hubble time, a very small coalescence rate that is
comparable to the rate by which massive galaxies merge.

It is possible that not all MBH coalescences contribute to the assembly of
the SMBH mass. In particular, MBHs that are born into tight binaries might
coalesce, even if they do not migrate into dense stellar systems or MBH
clusters.  If MBH binaries did coalesce without adding mass to the
assembled SMBHs, then the observed event rate would only provide an upper
limit on $f_{\rm MBH}$. MBHs that coalesced directly with SMBHs without
first undergoing coalescence with other MBHs would be invisible to Advanced
LIGO, but might be detected by the planned space mission {\it
LISA}\footnote{http://lisa.jpl.nasa.gov/} (Wyithe \& Loeb 2003).

\acknowledgements 

We thank Scott Hughes and Andrew Melatos for helpful comments. This work
was supported in part by NASA grant NAG 5-13292, and by NSF grants
AST-0071019, AST-0204514 for AL.

\end{document}